\documentclass{llncs}
\usepackage[latin1]{inputenc}
\usepackage[detect-all]{siunitx}
\usepackage{amsfonts}
\usepackage{amsmath}
\usepackage{amssymb}
\usepackage{graphicx}
\usepackage[ruled,longend]{algorithm2e}
\usepackage{stmaryrd}
\usepackage{fancybox}
\usepackage{url}
\usepackage{cite}
\usepackage{multirow}
\usepackage{subfig}
\usepackage{booktabs}

\newcommand{\strings}{\ensuremath{\mathcal{S}}}
\newcommand{\lookup}{\ensuremath{\text{Lookup}}}
\newcommand{\access}{\ensuremath{\text{Access}}}
\newcommand{\str}[1]{\ensuremath{\texttt{#1}}}

\newcommand{\rank}{\ensuremath{\textit{rank}}}
\newcommand{\select}{\ensuremath{\textit{select}}}

\newcommand{\newterm}[1]{\textit{#1}}

\newcommand{\abs}[1]{\ensuremath{\left\vert#1\right\vert}}

\newcommand{\Cpp}{C\raisebox{2pt}{{\relsize{-3}++}}\ }

\title{LZ-Compressed String Dictionaries}
\author{Julian Arz \and Johannes Fischer}
\institute{
  KIT, Karlsruhe, Germany.
  \email{julian.arz@student.kit.edu,johannes.fischer@kit.edu}
}

\begin{document}
\maketitle

\begin{abstract}
We show how to compress string dictionaries using the Lempel-Ziv
(LZ78) data compression algorithm.  Our approach is validated
experimentally on dictionaries of up to 1.5 GB of uncompressed text.
We achieve compression ratios often outperforming the existing
alternatives, especially on dictionaries containing many repeated
substrings. Our query times remain competitive.
\end{abstract}

\section{Introduction}
A \newterm{string dictionary} is a data structure that stores a set of words and identifies each word with a unique identifier. It has to support two straightforward operations: return a word, given its ID (\emph{access}), and return the ID of a given word (\emph{lookup}). Other operations, such as searching for all words with a certain suffix or prefix or containing some substring, are optional. String dictionaries are a basic tool for the processing and indexing of strings, whenever a mapping from a set of words to a unique ID is needed.
%Their applications are not restricted to domains of computer science, but extend to other sciences as well.

%In Molecular Biology, for example, string algorithms are used to analyze the structure of biological molecules. These can be seen as sequences of amino acids for proteins, or nucleotide bases for DNA strands, which are represented as strings over an alphabet of size 20 or 4, respectively. One of the most popular algorithms, BLAST \cite{Altschul90-blast}, is used to find similar occurrences of a given query sequence in a library of target sequences.%TODO cite GenBank
%The query string is parsed into substrings of a certain length, which are then searched in the library. Here, string dictionaries can be used to index the sequence database.

Though being a % nicht 'an', da Aussprache 'jubi...'
ubiquitous problem, to the best of our knowledge not much emphasis has been put on the \emph{size} of string dictionaries. A likely explanation for this is that in the past the set of strings to be stored in a dictionary were not big enough to justify complex data structures (the baseline algorithms were sufficient). For example, a typical application for string dictionaries are geographical information systems, where all geographic names of a region need to be stored in a dictionary. But even on a continental-size map the size of such a dictionary is manageable with traditional techniques. However, this situation has changed in recent years, as string dictionaries are becoming larger and larger, and also because on mobile devices (e.g.~GPS's), available memory is still a critical factor.

A rather classical application arises in information retrieval. Huge amounts of data have to be organized in a way that facilitates the extraction of small fragments. In document retrieval systems for example, collections of text documents are indexed so that a user can \newterm{query} them, say find all documents where a certain word occurs. Here, string dictionaries are used in an extended form: \newterm{inverted indexes} assign to every word present in the collection a list of documents they appear in, and a first step for locating a given term is to search it in the dictionary. These dictionaries can become quite big. While the total number of words in the English language is estimated to about 1\,000\,000 \cite{MA11-qacumdb}, recent crawls of web pages written in 10 different languages resulted in a data set of 200~million different words. This can be due to typing errors, but also stems from the fact that in some languages, for example in German, new words can be built by the concatenation of two existing words.

In databases of web platforms (e.g. social networks), string attributes are commonly used to store data such as user information, private messages or guest-book entries. Tables can consist of several string attributes and a unique primary key (ID), a number. This ID is usually stored as a foreign key in another table. A user is then able to either search for a string to use its ID in another context, or obtain the string to a given ID, which makes databases an ideal application for string dictionaries. Often, database columns are indexed to hasten the search, at the expense of additional space overhead.
%Using a compressed string dictionary, the space consumption of the table could be lessened, while preserving the performance improvement of the index.
\emph{Column based internal memory data bases} are another natural example where string dictionaries arise and are a hot research topic.

All of the given examples could profit from a reduction of the space overhead of the used string dictionary. Thus, the question arises how string dictionaries can be compressed. Two trivial solutions come to mind: First, we could regard a set of separate words concatenated to one string, and compress this string using any state of the art lossless data compression technique, e.g. \textit{deflate} or other algorithms.
% from the Lempel-Ziv-family. 
This results in great compression ratios, but poor efficiency for the required operations, as in the worst case the whole text has to be decompressed to access one word. Second, we could compress each word separately and use typical string dictionary methods like hashing to provide the functionality. In that case, the two operations were almost as fast as in traditional string dictionaries, but the compression is far from optimal.

%Both solutions do not satisfy the need for good compression rates \emph{and} fast operations. Thus there is a need to develop new data structures, \emph{compressed string dictionaries}, which do combine those two features.
%Namely, one wants compression ratios as high as those achieved when compressing the concatenated string, while preserving the query times of traditional string dictionaries. Of course, this seems unlikely, so convenient trade-offs have to be found. A method which has been widely used is practice is \emph{front coding}. For a faster access, a front coded dictionary can be represented as a \emph{trie}, but a traditional trie implementation increases the space overhead. A huge amount of work has been done on improving trie representation, and recently a space saving trie representation \cite{GO12-CompressedTriesPathDecomposition} which also supports fast access in practice has been proposed.

\subsection{Related Work}
\label{sect:related}
The immediate solution to the string dictionary problem is to store the strings in a \emph{trie}, where the leaves are annotated with the identifiers.
%Then the lookup operation becomes trivial. To support the access operation, we can augment the trie with \emph{backwards} pointers. Then going to the $i$'th leaf and following the backwards pointers towards the root recovers the $i$'th string in reverse order.
A related but more practical idea is to sort the strings lexicographically and encode a string as a pair $(\ell, \alpha)$, where $\ell$ is the length of the longest common prefix with its lexicographic predecessor and $\alpha$ is the remaining suffix. This idea is known as \emph{front coding}. To support fast access- and lookup-operations, every $k$'th string is stored \emph{verbatim}, for some suitably chosen value of $k$. However, none of these simple methods provides general-purpose compression.

Research on compressed string dictionaries is more recent, and we are aware of only a few works tackling this problem. The first is the \emph{compressed permuterm index} \cite{ferragina10compressed} that builds on the Burrows-Wheeler transformation. It supports a rich set of operations for IR tasks, but if restricted to our simple access/lookup functionality its space is not competitive. Brisaboa et al.~\cite{brisaboa11compressed} evaluate the practical performance of techniques like Huffman coding, hashing, front coding, grammar-based compression, and full text indexing. In brief, they find that (a) front coding with Hu-Tucker character compression and (b) Re-Pair-based indices provide the best time/space trade-offs.
The most recent work is due to Grossi and Ottaviano \cite{grossi12fast} and builds on previous ideas of the first author \cite{ferragina08searching}. It augments the basic trie idea with \emph{path decomposition}, and is shown to often perform better than \cite{brisaboa11compressed}. All approaches employ engineered implementations of \emph{succinct data structures} to achieve good practical performance.

\subsection{Our Contribution}
We improve the empirical performance of dictionary-based compressed string dictionaries, namely Lempel-Ziv (LZ) compressed string dictionaries. The only existing dictionary-based approach is Re-Pair \cite{brisaboa11compressed}. Compared to the latter, we achieve very competitive compression rates, often better. Accesses are 2--4 times slower, but lookups are up to twice as fast. Construction of our data structure is one order of magnitude \emph{faster} than Re-Pair. Compared to the path-decomposed tries \cite{grossi12fast}, we achieve about the same compression ratios, but often slower operations. A notable exception is a data set containing many often repeated substrings, where we can compress to about a third of the original file size, whereas the path-decomposed trie achieves only 1/2. In total, our approach proves to be very robust due to the direct use of a general-purpose compressor from the LZ-family.

\section{Preliminaries}
\label{sect:prelims}
In this section we introduce known concepts and tools on which our data structure is based. We start by formally defining our problem as follows. Let $\strings = \{s_0,\dotsc,s_{m-1}\} \subset \Sigma^\star$ be a set of $m$ strings, and let $n=\sum_{i<m}|s_i|$ denote their combined length. A \emph{string dictionary} over $\strings$ is a data structure that supports the following operations:
\begin{itemize}
\item $\lookup(s)$: return $-1$ if $s\not\in\strings$ or a unique identifier in $[0,m)$ otherwise.
\item $\access(i)$: get the string with identifier $i\in[0,m)$, $\lookup(\access(i))=i$.
\end{itemize}

\subsection{LZ78 Data Compression}
\label{sect:lz}
Our new algorithm is based on the LZ78 compression algorithm \cite{ziv78compression} for a string $S[0,n)$, which we are going to describe next. The algorithm proceeds by parsing $S$ from left to right and dividing it into blocks (called \emph{phrases}) that are one-letter extensions of previously seen phrases. The set of current phrases is called the \emph{phrase dictionary}. At the beginning of the algorithm the phrase dictionary contains only the empty string. Now assume that we have already parsed a prefix $S[0,i)$ of $S$, and that the phrase dictionary is $\mathcal{D}$. Then the next phrase is chosen to be $S[i,j]$ such that $S[i,j)$ is the longest string already in $\mathcal{D}$. Further, the new phrase $S[i,j]$ is added to $\mathcal{D}$. Due to the construction, the dictionary $\mathcal{D}$ is prefix-closed and is naturally represented with a trie.

\subsection{Succinct Data Structures}
\label{sect:succinct}

Consider a \emph{bit-string} $S[0,n)$ of
length $n$. We define the fundamental \emph{rank}- and \emph{select}-operations on $S$ as follows:
$\rank_1(S,i)$ gives the number of 1's in the prefix $S[0,i]$, and $\select_1(S,i)$ gives the
position of the $i$'th 1 in $S$, reading $S$ from left to right ($0 \le i < n$). Operations
$\rank_0(S,i)$ and $\select_0(S,i)$ are defined similarly for 0-bits. The following lemma summarizes
a by-now classic result; well-performing practical implementations of this lemma exist, we used the SDS-Library \cite{sdsl}.
\begin{lemma}[see, e.g., \cite{munro01succinct}]
\label{lem:fid}
A bit-string of length $n$ can be represented in $n+o(n)$ bits such that rank- and select-operations are supported in $O(1)$ time.
\end{lemma}

\section{New Data Structure}
\label{sect:theoretical}

We now present the theory of our new algorithm. We proceed by first showing a data structure that supports the access-operation (Sect.\ \ref{sect:first}), and then modifying this data structure to also support the lookup-operation efficiently (Sect.\ \ref{sect:second}).

\subsection{Basic Idea: Supporting Access}
\label{sect:first}

We explain a first idea how to adapt the LZ78-parsing from Sect.~\ref{sect:lz} to the string dictionary problem. For ease of explanation, we assume that each string $s_i$ is terminated by a unique letter $\#_i\not\in\Sigma$. Then we can \emph{concatenate} all strings into a single large string $S=s_0s_1\dots s_{m-1}$ of length $n$, without losing information about the word boundaries. The basic approach is to compress $S$ with the LZ78-parsing algorithm from Sect.~\ref{sect:lz} and store the resulting LZ-trie. Note that due to the unique separators $\#_i$, there is a phrase ending at the end of every string, and hence every string also starts with a new phrase.

\begin{figure}[t]
  \centering
  \includegraphics{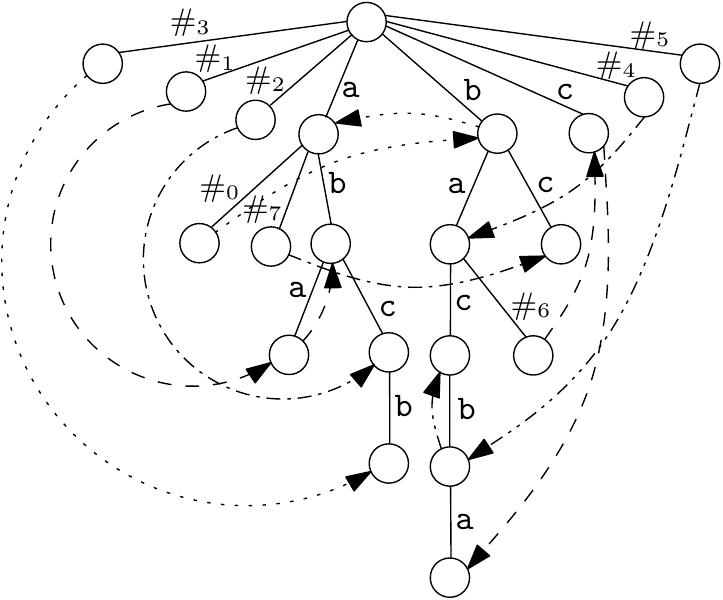}
  \caption{LZ-trie for the set of strings $\strings=\{\str{a}|\str{b}|\str{a}|$, $\str{ab}|\str{aba}|$, $\str{abc}|$, $\str{abcb}|$, $\str{ba}|$, $\str{bac}|\str{bacb}|$, $\str{bacba}|\str{c}|\str{ba}|$, $\str{bc}|\str{a}|\}$, with the parsing indicated by ``$|$''. All non-solid lines are predecessor links enabling the efficient access to strings.}
  \label{fig:example1}
\end{figure}

For the recovery of the original strings in \strings, we \emph{link} the phrases from one string as follows. Suppose $s_i$ is parsed as $s_i=p^i_0|p^i_1|\dots |p^i_{k-1}|$, where the $p^i_j$ are phrases. Note that each $p^i_j$ corresponds to a unique node $v^i_j$ in the LZ-trie. For $1\le j < k$, we make a link from $v^i_j$ to $v^i_{j-1}$. We call those links the \emph{predecessor links}. See Fig.\ \ref{fig:example1} for an example. We can store an additional array $A[0,m)$ such that $A[i]$ points to the node corresponding to the last phrase $p^i_{k-1}$ of $s_i$ (the one ending with $\#_i$).

Now the access-operation can be easily supported. Suppose we want to answer $\access(i)$, and that the parsing of $s_i$ is $p^i_0|p^i_1|\dots |p^i_{k-1}|$. Then we first go to node $v^i_{k-1}=A[i]$ and recover $s_i$'s last phrase $p^i_{k-1}$ by following the path from $v^i_{k-1}$ towards the root.
%(the LZ-trie also stores backwards pointers).
Then we follow the predecessor link of $v^i_{k-1}$ to $v^i_{k-2}$ and recover the penultimate phrase. This goes on until we have recovered the first phrase $p^i_0$, which happens iff the predecessor link is \texttt{nil}. As a result, we have recovered the $i$'th string from right to left.

Unfortunately, this data structure does not readily support the lookup in optimal $O(|s|)$ time, because the parsing of phrases is not unique. For example, consider querying for the string \str{aba} in the example of Fig.\ \ref{fig:example1}. Matching it greedily in the LZ-trie we arrive at the node spelling the string \str{aba}, from which we cannot derive a correct identifier. Indeed, what we should have done is matching \str{aba} as it was parsed: first \str{a}, then \str{b}, and finally $\str{a}\#_1$, from which we could have derived that '1' is the true identifier of the string \str{aba}. However, the LZ-trie does not seem to contain sufficient information to decide that after matching the first \str{a} we should have started a new phrase.

\subsection{Modification for Supporting Lookup}
\label{sect:second}
The problem of the previous section is that the parsing of a string $s_i$ depends on the past, i.e., on the parsing of all strings $s_j$ for $j<i$.
We resolve this problem by \emph{reparsing} the strings to make their parsing unique and enable a greedy parsing of query strings. More precisely, we first construct the LZ-trie from Sect.~\ref{sect:first} (excluding the predecessor links and end-of-string markers $\#_i$). Then, we run through the dictionary $\strings$ \emph{again} and reparse all strings by matching them \emph{greedily} in the existing trie. This implies that phrases can now be used multiple times. The advantage is that the parsing is now \emph{unique} in the following sense: say that $s_i$ is parsed as $s_i=p^i_0|p^i_1|\dots |p^i_{k-1}|$, and that a different string $s_j=p^j_0|p^j_1|\dots |p^j_{\ell-1}|$ is prefixed by $p^i_0p^i_1\dots p^i_{y-1}$ for some $y\le k$. Then $p^i_x = p^j_x$ for all $x<y$, and further, if the longest common prefix between $s_i$ and $s_j$ extends $r$ characters into $p^i_y$, then those $r$ characters are also a prefix of $p^j_y$.

\begin{figure}[t]
  \centering
  \subfloat[The reparsed LZ-trie.]{\label{fig:example2}\includegraphics{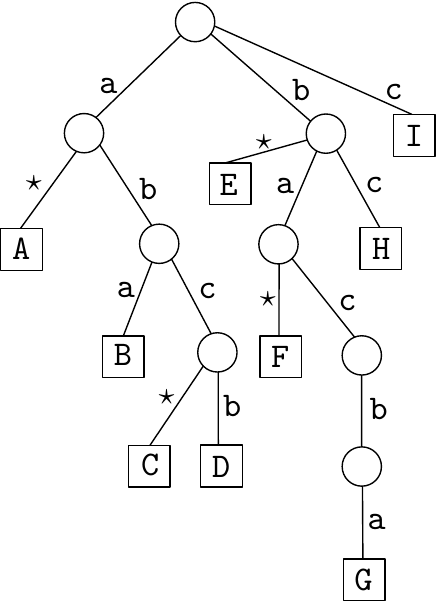}}
  \hfill
  \subfloat[The resulting phrase trie.]{\label{fig:example3}\includegraphics{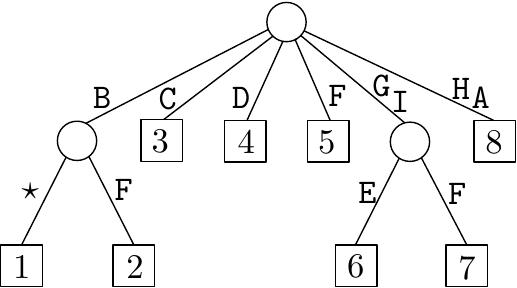}}
  \caption{The final data structures. The set of strings from Fig.~\ref{fig:example1} is now parsed as $\strings=\{\str{aba}|$, $\str{aba}|\str{ba}|$, $\str{abc}|$, $\str{abcb}|$, $\str{ba}|$, $\str{bacba}|\str{c}|\str{b}|$, $\str{bacba}|\str{c}|\str{ba}|$, $\str{bc}|\str{a}|\}$, which corresponds to the strings $\strings'=\{\str{B}$, $\str{BF}$, $\str{C}$, $\str{D}$, $\str{F}$, $\str{GIE}$, $\str{GIF}$, $\str{HA}\}$ in the phrase alphabet.}
  \label{fig:final}
\end{figure}

For example, string $\str{bacbacb}$ in Fig.~\ref{fig:example1} is now parsed $\str{bacba}|\str{c}|\str{b}$ instead of $\str{bac}|\str{bacb}$. Although the new parsing is longer in this case, it is parsed similar to the next string $\str{bacba}|\str{c}|\str{ba}$. See Fig.~\ref{fig:example2} for the resulting LZ-trie, where phrases that are a prefix of a different phrase are terminated with a special character ``$\star$'' in order to make all phrases end at a leaf of the LZ-trie. (This allows us to identify all phrases with leaf identifiers; in Fig.~\ref{fig:example2} denoted by upper case letters.)

Two issues arise that have to be dealt with now:
\begin{itemize}
\item Some nodes from the original LZ-trie could now be superfluous, since they are not reached anymore by any used phrase. Such nodes can simply be deleted from the resulting trie.
\item More seriously, it could happen that the greedy parsing cannot continue, for example when a phrase is parsed longer than originally, but there is no outgoing edge from the root with the next character. This problem can be solved by initially inserting all single letters $a\in\Sigma$ into the trie.
\end{itemize}

Due to the multiple use of phrases we cannot work with plain predecessor pointers as before. To overcome this, we construct \emph{another} trie consisting of the parsed phrases in the new \emph{phrase alphabet}. We call this second trie the \emph{phrase trie}. See Fig.~\ref{fig:example3} for an example. (This trie contains exactly the predecessor pointers, but arranged in a form more suitable for querying, as we shall see.)

Now $\access(i)$ works slightly differently: jump to the $i$'th leaf of the phrase trie and find the phrases of $s_i$. Those phrases can now be easily recovered using the LZ-trie, since there they correspond to leaves. The time is the optimal $O(|s_i|)$.

The advantage of the new structure is that it also enables optimal $\lookup(s)$: first, parse $s$ greedily using the LZ-trie, e.g.~$s=p_0|p_1|\dots|p_{k-1}|$. Then we try matching the parsed phrases in the phrase trie; this takes $O(k)$ time (assuming perfect hashing). If matching is successful and ends in a leaf, we return the identifier stored there (the string ID). Otherwise $s$ does not occur in $\strings$. The total time for this process is optimal $O(|s|)$.

\section{Implementation Details}
\label{sect:implementation}

In this section we describe our implementation that builds on the data structure from Sect.~\ref{sect:theoretical}. As often in algorithmic engineering, we sometimes deviate from the theoretical proposal and sacrifice the optimal running times in exchange for a faster practical performance.

\subsection{Representation of the LZ-Trie}
The choice of the LZ-trie implementation offers a trade-off between time for each of the two operations and space overhead. Besides a trie, other data structures are possible as well: the required operations are access and \verb|longest_prefix| (finding the longest element which is a prefix of a given string).

We present two representations. The first is a well-tuned existing trie implementation, the path-decomposed tries of Grossi and Ottaviano \cite{grossi12fast}. One thing to remark is that while the path decomposition has the advantage of fast operation times due to high cache locality, its drawback is that in its original form, only the leaves of the original trie can be accessed, whereas the inner nodes are ``hidden'' in the path decomposition.
Instead of
appending a unique character ``$\star$'' to the end of each phrase as in Sect.~\ref{sect:second},
we identify a phrase with a tuple consisting of the subpath it ends on in the path decomposition and the offset, counting from the beginning of this subpath. We map these tuples to ordinary numbers using two bit vectors enhanced with rank/select-support.
We found this to be more space efficient than the explicit end-of-string character.

The second representation is based on a front coding dictionary, as the one described in Sect.~\ref{sect:related}. We support \verb|longest_prefix| using a characteristic of our LZ-parsing: when a parsed string $s$ and a phrase $p$ have an lcp $r$, the longest prefix is at least as long as $r$. We search the given string in the dictionary, first with a binary search on the explicitly stored entries and then with a linear search in a bucket. If this search does not find prefix, we either search the lcp of that bucket's first entry and the string, or abort the search (and return $-1$) based on the already compared strings. Thus, at most two search operations are performed.

\subsection{Representation of the Phrase Trie}
\label{sect:phrase_trie}
Due to the potentially very large alphabet ($p$, the number of different phrases, can be very high), existing trie implementations cannot be used efficiently for storing the phrase trie. Instead, we chose a different approach, as explained next.

\begin{figure}[t]
  \centering
  \subfloat[Normal.]{\label{fig:example4}\includegraphics{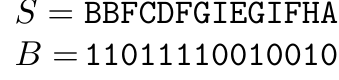}}
  \hfill
  \subfloat[Sorted by length.]{\label{fig:example5}\includegraphics{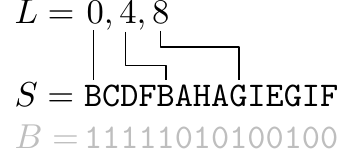}}
  \hfill
  \subfloat[First phrases omitted.]{\label{fig:example6}\includegraphics{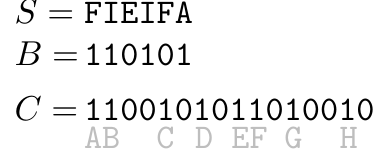}}
  \caption{The linearized phrase trie, where (b) and (c) show two optimization techniques. Information in gray need not be stored explicitly.}
  \label{fig:final2}
\end{figure}

We linearize the phrase trie by juxtaposing all parsed phrases into a string $S$, see Fig.~\ref{fig:example4} for the running example. In order to know where a string $s_i$ starts, we store an additional bit vector $B$ of the same length as $S$, where a '1' indicates the beginning of a string. Then the parsing of the $i$'th string can be found by $\select_1(B,i)$, and hence the access-operation can be easily supported. On the other hand, for $\lookup(s)$ we now have to search the parsing of $s=p_0|p_1|\dots|p_{k-1}|$ in $S$; we do this by a \emph{binary search} with the help of $\select$-operations over $B$. As an example, consider searching $s=\str{bc}|\str{ba}|\str{a}|$, which corresponds to $s'=\str{HFA}$ in the phrase alphabet. Since there are $n=8$ strings in $\strings$, we first go to the 4th (middle) string by $\select_1(B,4)$, see that it is $\str{D}$, and since it is alphabetically smaller than the parsed query string $\str{HFA}$, we continue the search in the right half. To make the binary search work, the parsed strings are sorted before writing them to $S$.
%Of course, this binary search assumes that the parsed strings are in sorted order, which must be done at this point if it has not been done before.

Two optimizations can be applied to this base variant of the linearized phrase trie. The first is to rearrange the parsed strings in $S$ such that they are first sorted by their length (number of phrases), and then lexicographically. %Then the bit-vector $B$ becomes superfluous and can hence be discarded, 
The bit vector $B$ can then be discarded, since the string beginnings can be calculated arithmetically within a range of strings of equal length (see Fig.~\ref{fig:example5}). We need an additional small array $L$ to know where the phrases of a given length begin in $S$. The access-operation uses this array to find the range containing the requested index. For the lookup-operation, we can identify the correct range directly by counting the number of phrases the searched string was parsed into. Thus, during the binary search we omit the $\select$-operations which, although constant-time, introduce a considerable time overhead due to several table lookups.

The goal of this optimization is thus not so much a reduction in space (as $B$ is small compared to $S$) but rather an acceleration of the lookup-operation. In our experiments we found that most of the strings are parsed into two or three phrases (mean $\approx 2.7$), but a small percentage is parsed into more (up to the hundreds) of phrases. To account for this, we chose to sort by length only the strings with a small number of phrases (at most 5), and handle all larger strings as before (\emph{with} the bit vector $B$, which is now very short). The additional time overhead for the access-operation is then negligible.

The second optimization is to \emph{omit} the first phrase from every string in a lexicographically sorted range of $S$. Then we need an additional bit vector $C$ that encodes how many strings start with a given character in the phrase alphabet. This can be done, e.g., by writing a 1 for every phrase character, followed by $k$ 0's if there are $k$ strings starting with that phrase (see Fig.~\ref{fig:example6}). Preparing $C$ for select queries (on both 0- and 1-bits) then allows to recover the original contents of $S$. For $\access(i)$, we count the number of 1-bits up to the $i$'th 0-bit in $C$ to retrieve the first phrase. For the remaining phrases, we proceed similarly for $B$. During the lookup-operation, we only need to search in the range of parsings starting with $p_0$. We find this range with a $\select_1(C, p_0)$. 

This optimization turned out to be very effective for large dictionaries, since with $p$ phrases we save $n\lg p$ bits by dropping the first phrases, whereas array $C$ occupies only $p+n$ bits, much less than $n\lg p$ for large $n$ and typical values of $p=O(n/\lg n)$. The lookup-operation is faster as well, because less comparisons are performed during the binary search.

Both optimizations can be combined. Then for each range of strings with equal parsing length $\ell$ a bit vector $C_{\ell}$ is used. Again we only support the ranges up to a certain number of phrases. If there are $n_{\ell}$ strings in a range, the size of $C_{\ell}$ is $n_{\ell} + p$ bits, opposed to the savings of $\ell n_{\ell}$ bits for omitting $B$ in that range. Thus this combination yields better compression rates only if the number of strings in one range is large enough to compensate for the $p$ bits added for every supported range.

\section{Experimental Results}
We perform several experiments on real world data as well as a synthetic data set to evaluate our algorithm. We also compare our data structure to other relevant structures for compressed string dictionaries.
\paragraph{Setting.}
We use the following data sets, of which URLs were also used in two previous experiments \cite{grossi12fast,brisaboa11compressed}, and Wiki was used only by Grossi and Ottaviano \cite{grossi12fast}:
\begin{description}
\item[Wiki] consists of all page titles of the English Wikipedia of April 2011,
\item[URLs] are the URLs of a 2002 crawl by the UbiCrawler \cite{BCSV04-UbiCrawler} on the \texttt{.uk} domain,
\item[DNA] is the DNA data set from Pizza\&Chili Corpus \cite{PizzaChiliDNA}, split into strings of 30 characters, and
\item[synth-$\boldsymbol{\alpha\beta\alpha}$] is a data set we constructed artificially. These are strings of the form $\alpha_1\beta\alpha_2$, where the $\alpha_i$'s are randomly chosen strings which occur multiple times. Strings $\beta$ are short strings to separate these blocks. See Appendix~\ref{sect:syntheticDatasets} for technical details on 
%the constitution of 
this data set.
This data set was chosen to show that path decomposition does not always result in better compression ratios than purely dictionary-based methods.
\end{description}

Our testing machine is an AMD Opteron 8350, equipped with 4 cores (of which we only use one) and 64 GiB of main memory. The CPU is clocked with 2.0 GHz and is supported by 2 MiB Cache. The machine is running Ubuntu 10.04.4 LTS (kernel 2.6.32). All algorithms were implemented in \Cpp and compiled using the GNU \Cpp compiler version 4.4.3 with optimization level -O3. 
We chose 1,000,000 random indices and strings from each data set for the timing of the operations.
All measured times are averaged over 3 runs.

\paragraph{Reparsing.}
\begin{table}[t]
\caption{Comparison of the size of LZ-parse before and after the reparsing. All values are $10^3$.}
\label{tab:datasets}
\centering
\begin{tabular}{lrrrrrrrr}
\toprule
& \multicolumn{2}{c}{synth-$\alpha\beta\alpha$} & \multicolumn{2}{c}{Wiki} & \multicolumn{2}{c}{URLs}  & \multicolumn{2}{c}{DNA}\\
\cmidrule(lr){2-3} \cmidrule(lr){4-5} \cmidrule(lr){6-7} \cmidrule(lr){8-9}
		 & before & after & before & after 	& before & after & before & after  \\
\midrule
\#Nodes 		& 24340 & 15722 & 13597 & 12536	& 46716	& 34032 & 24832	& 24054 \\
\#Phrases ($p$) & 24340 & 6873 & 13597  & 8624	& 46716	& 17652	& 24832	& 18105 \\
\#Parsing		 & 24340 & 21324 & 13597& 19782 & 46716	& 55008	& 24832	& 37123 \\
\bottomrule
\end{tabular}
\end{table}
Table~\ref{tab:datasets} gives numbers on the size of the LZ-trie before and after the reparsing. \#Nodes is the number of nodes in the (uncompacted) trie. The smaller numbers after reparsing are a result of cutting off entire subtrees from the LZ-trie. The second row gives the number of nodes representing phrases (lower after reparsing since not every prefix of a phrase is necessarily also a phrase). The row ``\#Parsing'' gives the size of the parsing.
% (which does not affect the size of the trie).
All these numbers are equal before the reparsing as every node in the trie corresponds to a phrase and every phrase is used exactly once in the parsing.
We observe that the reparsing reduces the size of the LZ-trie by about 27\% for URLs and 8\% for Wiki. For our synthetic data set, the reduction is 35\%. These results confirm the intuition that the LZ-based reparsing strategy excels for collections of strings where sufficiently long substrings occur multiple times, e.g.    ``$\str{http://}$'', ``$\str{index}$'' or ``$\str{.html}$'' for URLs. The size of the parsing is increased by between 18\% and 50\%, but this change is not reflected in the size of the trie, and it is the role of our phrase trie implementation to cope with this increase.

\begin{table}[t]
\caption{The compression ratio of our data structure broken down into its two components. All percentages are in relation to the original file size.}
\label{tab:distribution}
\centering
\subfloat[][LZ-Trie represented by PDT (LZT-pd)]{
\centering
\begin{tabular}{lrrrr}
\toprule
			& synth-$\alpha\beta\alpha$ & Wiki & URLs & DNA	\\
File Size [\si{\mega\byte}] & 212 & 171 & 1\,439 & 400	\\
\midrule                                                                                                                    
LZ-Trie [\%] 				& 11.1 & 11.6 & 3.3 & 8.6  	\\
Phrase Trie [\%]  			& 23.8 & 23.0 & 8.8 & 21.4 	\\ [3pt]
Combined [\%]  				& 34.9 & 34.6 & 12.2 & 30.0	\\
\bottomrule
\end{tabular}}
\subfloat[][by front coding (LZT-fc)]{
\centering
\begin{tabular}{rrrr}
\toprule
synth-$\alpha\beta\alpha$ & Wiki & URLs & DNA\\
212 & 171 & 1\,439 & 400	\\
\midrule                                                                                                                    
15.8 & 20.5 & 7.7 & 19.1  	\\
23.8 & 23.0 & 8.8 & 21.4 	\\ [3pt]
39.6 & 43.4 & 16.4 & 40.5	\\
\bottomrule
\end{tabular}}
\end{table}
\paragraph{Other Data Structures.}
We compare our data structure to previously known techniques.
We did not include uncompressed dictionaries (like TX-trie \cite{tx}) in our evaluation, as they were already shown to use much more space in previous studies \cite{brisaboa11compressed,grossi12fast}.
LZT is our approach of the reparsed LZ-trie, using the implementation described in Sect.~\ref{sect:implementation}. We examine the trade-offs for two different representations of the LZ-trie, one using path decomposition (LZT-pd) and the other using front coding with bucket size 16 (LZT-fc).
%We use only the second optimization of Sect.~\ref{sect:phrase_trie} as it achieves the lowest space overhead. It reduces the size of the phrase trie representation by 21--37\% compared to the base variant. Combining it with the partitioning into ranges of equal size turned out not to be beneficial for the tested instances.
Table~\ref{tab:distribution} gives more information about the distribution of space consumption on our two components. In all cases, the phrase trie representation accounts for about than two thirds of the overall space (slightly more than two thirds for LZT-pd, and slightly less for LZT-fc).

Tables~\ref{tab:experiments_synth} to \ref{tab:experiments_DNA} present the experimental comparison with other approaches.
PDT is the centroid path-decomposed trie \cite{grossi12fast} in the compressed variant. The comparison between LZT and PDT is interesting, as we actually use the PDT in our implementation. There is a remarkable duality between the two: PDT uses a grammar-based compression on a (linearized) trie built on all strings, while LZT builds a trie on a grammar-based compression scheme.
Re-Pair, front coding (FC) and Hu-Tucker front coding (HTFC) are examined in \cite{brisaboa11compressed}.
We experimentally deduced the best bucket size to be 8 for both FC and HTFC, slightly favoring the compression.

\begin{table}[t]
\sisetup{
table-format=2.2,
table-number-alignment=right
}
\caption{Comparison of our data structure with others.}
\centering
\subfloat[][synth-$\alpha\beta\alpha$ ($5.4\times10^6$ strings)]{
\label{tab:experiments_synth}
\begin{tabular}{lS[table-format=5.2]rSS}
\toprule
 & {constr} & {cmpr}  & {access}  & {lookup} \\
 & [\si{\second}] & [\%] & [\si{\micro\second\per ID}] & [\si{\micro\second\per str}] \\
\midrule
LZT-pd & 177.8 & 34.9 & 15.3 & 16.7 \\
LZT-fc & 96.0 & 39.6 & 5.4 & 12.8 \\[3pt]
Re-Pair & 2959.7 & 31.0 & 3.8 & 14.4 \\
PDT & 291.0 & 53.2 & 4.1 & 4.1 \\
FC & 0.55 & 68.0 & 0.59 & 2.0 \\
HTFC & 2.52 & 49.4 & 4.8 & 7.0 \\
\bottomrule
\end{tabular}}
\qquad
\subfloat[][Wiki ($8.5\times10^6$ strings)]{
\label{tab:experiments_Wiki}
\begin{tabular}{lS[table-format=5.2]rSS}
\toprule
 & {constr} & {cmpr}  & {access}  & {lookup} \\
 & [\si{\second}] & [\%] & [\si{\micro\second\per ID}] & [\si{\micro\second\per str}] \\
\midrule
LZT-pd & 99.0 & 34.6 & 9.9 & 10.7  \\
LZT-fc & 53.7 & 43.4 & 3.3 & 8.2 \\[3pt]
Re-Pair & 1017.6 & 41.5 & 3.8 & 14.4 \\
PDT & 85.3 & 32.1 & 4.1 & 4.1 \\
FC & 0.79 & 60.2 & 0.62 & 2.1 \\
HTFC & 2.35 & 43.2 & 2.6 & 5.0 \\
\bottomrule
\end{tabular}}

\subfloat[][URLs ($18.5\times10^6$ strings)]{
\label{tab:experiments_URLs}
\begin{tabular}{lS[table-format=5.2]rSS}
\toprule
 & {constr} & {cmpr}  & {access}  & {lookup} \\
 & [\si{\second}] & [\%] & [\si{\micro\second\per ID}] & [\si{\micro\second\per str}] \\
\midrule
LZT-pd & 311.9 & 12.2 & 15.8 & 16.7  \\
LZT-fc & 197.8 & 16.4 & 4.7 & 14.4 \\[3pt]
Re-Pair & 12069.7 & 12.4 & 4.2 & 31.0 \\
PDT & 244.2 & 13.6 & 6.3 & 6.2 \\
FC & 2.8 & 32.7 & 0.74 & 3.5 \\
HTFC & 9.9 & 24.4 & 5.7 & 9.9 \\
\bottomrule
\end{tabular}}
\qquad
\subfloat[][DNA ($12.9\times10^6$ strings)]{
\label{tab:experiments_DNA}
\begin{tabular}{lS[table-format=5.2]rSS}
\toprule
 & {constr} & {cmpr}  & {access}  & {lookup} \\
 & [\si{\second}] & [\%] & [\si{\micro\second\per ID}] & [\si{\micro\second\per str}] \\
\midrule
LZT-pd & 217.8 & 30.0 & 16.0 & 16.0  \\
LZT-fc & 120.1 & 40.5 & 3.9 & 11.5 \\[3pt]
Re-Pair & 15537.4 & 37.2 & 2.3 & 12.7 \\
PDT & 188.1 & 26.7 & 5.9 & 5.9 \\
FC & 1.3 & 69.7 & 0.59 & 2.3 \\
HTFC & 5.3 & 30.9 & 2.5 & 5.1 \\
\bottomrule
\end{tabular}}
\end{table}

We observe that the compression ratio of LZT-pd is among the lowest for all data sets and is at most 12 \% higher than the best for each set. Remarkably, it achieves the best compression ratio for the URLs data set.
The nature of this set (long common prefixes) should intuitively favor the front coding--based data structures, but they do not detect the common substrings which are not at the beginning of the strings.
For our synthetic data set the difference to PDT is even more pronounced, as our data structure compresses to 34.0\%, whereas PDT achieves only 53.2\% compression rate. Re-Pair is even slightly better in this case, but the construction is more than 15 times longer.
As expected, the compression comes at the cost of higher times for the operations. This is because the strings are parsed into an average of 2.3--4 phrases (depending on the data set), and therefore this amount of elementary trie operations has to be performed, while PDT only requires one such operation.
The variant LZT-fc alleviates this deficiency by sacrificing slightly more space in exchange for much faster access-times. Lookup-times are also sped up (though not as significantly, but still faster than Re-Pair).

\section{Conclusions}
We proposed a new LZ-like parsing for string dictionaries that allows greedy trie-based pattern matching. In our implementation we combined this parsing with space-efficient representation techniques for the resulting two tries.
%, most notably the substitution of the first phrase of every parsed string with a bit vector.
Our data structure competes with other data structures regarding query times, and it often achieves better compression ratios, particularly for string collections containing highly repetitive patterns. 
Our work was not focused on the trie implementation. Representations better suited for phrases of an LZ-parse can achieve even better results. We also aim to investigate other parsings and grammars which might have convenient characteristics.

\section*{Acknowledgments}
We thank Giuseppe Ottaviano for providing his data sets, and Francisco Claude and Miguel \'Angel Mart\'inez for the source codes of their implementations. Further thanks go to Pawe\l{} Gawrychowski for interesting discussions on this topic.

\bibliographystyle{abbrv}
\bibliography{paper}
%\newpage
\appendix
\section{Synthetic Data Sets}
\label{sect:syntheticDatasets}
\paragraph{synth-$\alpha\beta\alpha$.}
Let $\Sigma^\alpha = \left\lbrace a,\dotsc,z\right\rbrace$ and $\Sigma^\beta = \left\lbrace !,\dotsc,@\right\rbrace$. We have $\abs{\Sigma^\alpha} = 26$, $\abs{\Sigma^\beta} = 32$ and $\Sigma^\alpha \cap \Sigma^\beta = \emptyset$. 
We define two pools (a.k.a.~multisets) of substrings. Every element of a pool is only used once in a constructed string.
Let $\Gamma$ be a pool of randomly generated strings out of $\left(\Sigma^\alpha\right)^{16}$. Each string occurs 32 times. Another pool $\Phi$ is built upon all lexicographically ordered strings of length 6 over the alphabet $\Sigma^\beta$. There are $\binom{32}{6}$ of these strings. Each of these strings occurs 6 times, so the size of $\Phi$ is $6 \binom{32}{6} = 5\,437\,152$.
Then, synth-$\alpha\beta\alpha$ consists of $\abs{\Phi}$ strings of the form $\alpha_1\beta\alpha_2$, where the $\alpha_i$ are taken at random from $\Gamma$ and the $\beta$ from $\Phi$. Thus, $\Gamma$ has to be twice as large as $\Phi$, which means it has to contain $\abs{\Phi} / 16 = 339\,822$ different strings.

\end{document}